\documentclass[aps,prl,twocolumn,groupedaddress,longbibliography]{revtex4-1}

\usepackage{graphicx}

\begin{document}
	

\title{Electric Rydberg-atom interferometery}


\author{J. E. Palmer}
\author{S. D. Hogan}
\affiliation{Department of Physics and Astronomy, University College London, Gower Street, London WC1E 6BT, United Kingdom}
	

\date{\today}
	
\begin{abstract}
An electric analogue of the longitudinal Stern-Gerlach matter-wave interferometer has been realized for atoms in Rydberg states with high principal quantum number, $n$. The experiments were performed with He atoms prepared in coherent superpositions of the $n=55$ and $n=56$ circular Rydberg states in zero electric field by a $\pi/2$ pulse of resonant microwave radiation. These atoms were subjected to a pulsed inhomogeneous electric field to generate a superposition of momentum states before a $\pi$ pulse was applied to invert the internal states. The same pulsed inhomogeneous electric field was then reapplied for a second time to transform the motional states to have equal momenta before a further $\pi/2$ pulse was employed to interrogate the final Rydberg state populations. This Hahn-echo microwave pulse sequence, interspersed with a pair of equivalent inhomogeneous electric field pulses, yielded two spatially separated matter waves. Interferences between these matter waves were observed as oscillations in the final Rydberg state populations as the amplitude of the pulsed electric field gradients was adjusted. 
\end{abstract}
	
	
\maketitle
	

The use of inhomogeneous electric or magnetic fields to exert forces on neutral gas-phase atoms and molecules has been central to developments in atomic and molecular physics over the last century. These began with the demonstration of space quantization by Gerlach and Stern~\cite{gerlach22a,gerlach22b} and subsequently included, e.g., the measurement of nuclear magnetic moments~\cite{rabi38a} which led to the development of nuclear magnetic resonance spectroscopy (see, e.g.,~\cite{emsley07a}); the preparation of quantum-state-selected molecular beams~\cite{bennewitz55a} and the realization of the ammonia maser~\cite{gordon55a}; magnetic trapping cold atoms~\cite{migdall85a} which facilitated the formation of Bose Einstein condensates~\cite{cornell02a,ketterle02a}; multistage Stark~\cite{bethlem99a} and Zeeman~\cite{vanhaecke07a,narevicius07a} deceleration, which have been exploited, e.g., to study molecular scattering at low temperatures~\cite{gilijamse06a,vonZastrow14a,allmendinger16a}, and for precision spectroscopy~\cite{vanveldhoven04a, jansen15a,jansen18a}; and electric traps for neutral atoms and molecules~\cite{bethlem02a,hogan08a} that have allowed studies of excited state decay processes on timescales not achievable in beam experiments~\cite{gilijamse07a,vandeMeerakker05a,seiler11a,seiler16a}. 

The experiments of Gerlach and Stern also opened opportunities to exploit the forces exerted by inhomogeneous fields on samples in coherent superpositions of internal states with different dipole moments, to generate coherent superpositions of momentum states for matter-wave interferometry~\cite{bohm51a,wigner63a,schwinger88a}. This led to the realization of an early electric atom interferometer~\cite{sokolov70a,sokolov73a}, and later the magnetic longitudinal Stern-Gerlach interferometer~\cite{miniatura91a,miniatura92a,nicChormaic93a} -- both demonstrated with H atoms in metastable $n=2$ levels. More recently, related experiments have been performed with laser cooled ground-state Rb atoms using sequences of pulsed radio-frequency and inhomogeneous magnetic fields~\cite{machluf13a}. These Stern-Gerlach matter-wave interferometers are distinct from, and complementary to, light-pulse and grating interferometers~\cite{cronin09a,hornberger12a}.

Here we demonstrate for the first time an electric analogue of the longitudinal Stern-Gerlach matter-wave interferometer for atoms in Rydberg states. This device exploits the state-dependent forces exerted by inhomogeneous electric fields on samples in states with large static electric dipole moments. Such forces were first exploited in an analogue of the original experiment of Gerlach and Stern, but using Kr Rydberg atoms, by Townsend~\emph{et al.}~\cite{townsend01a}. This work subsequently led to the development of the methods of Rydberg-Stark deceleration for controlling the translational motion of atoms or molecules composed of matter and antimatter~\cite{hogan16a,cassidy18a}. In all Rydberg-Stark deceleration experiments up to now, samples were prepared in stationary Rydberg eigenstates by laser photoexcitation. Thus, the existing methodologies may be classified as incoherent Rydberg atom or molecule optics. Here we report the results of Rydberg-Stark deceleration experiments performed with coherent superpositions of Rydberg states, and in doing so open a new era of coherent Rydberg atom optics. The Rydberg-atom interferometry reported here is distinct from time-domain internal-state Ramsey interferometry~\cite{ramsey90a}, in that the interference patterns presented here reflect matter-wave (or de~Broglie-wave) interference between two spatially separated momentum components into which each individual Rydberg atom is split, and interferometry schemes used in microwave cavity quantum electrodynamics experiments with Rydberg atoms~\cite{raimond01a}. The methods and results presented provide new opportunities for studies of long-range Rydberg interactions, may find application in tests of the weak equivalence principle for neutral atoms composed of antimatter, e.g., Rydberg antihydrogen~\cite{kellerbauer08a,amole13a} or positronium~\cite{mills02a,cassidy14a}, and are of interest for studies of spatial decoherence and wavefunction collapse in macroscopic quantum systems~\cite{bassi13a}.

\begin{figure}
\begin{center}
\includegraphics[width = 0.45\textwidth, angle = 0, clip=]{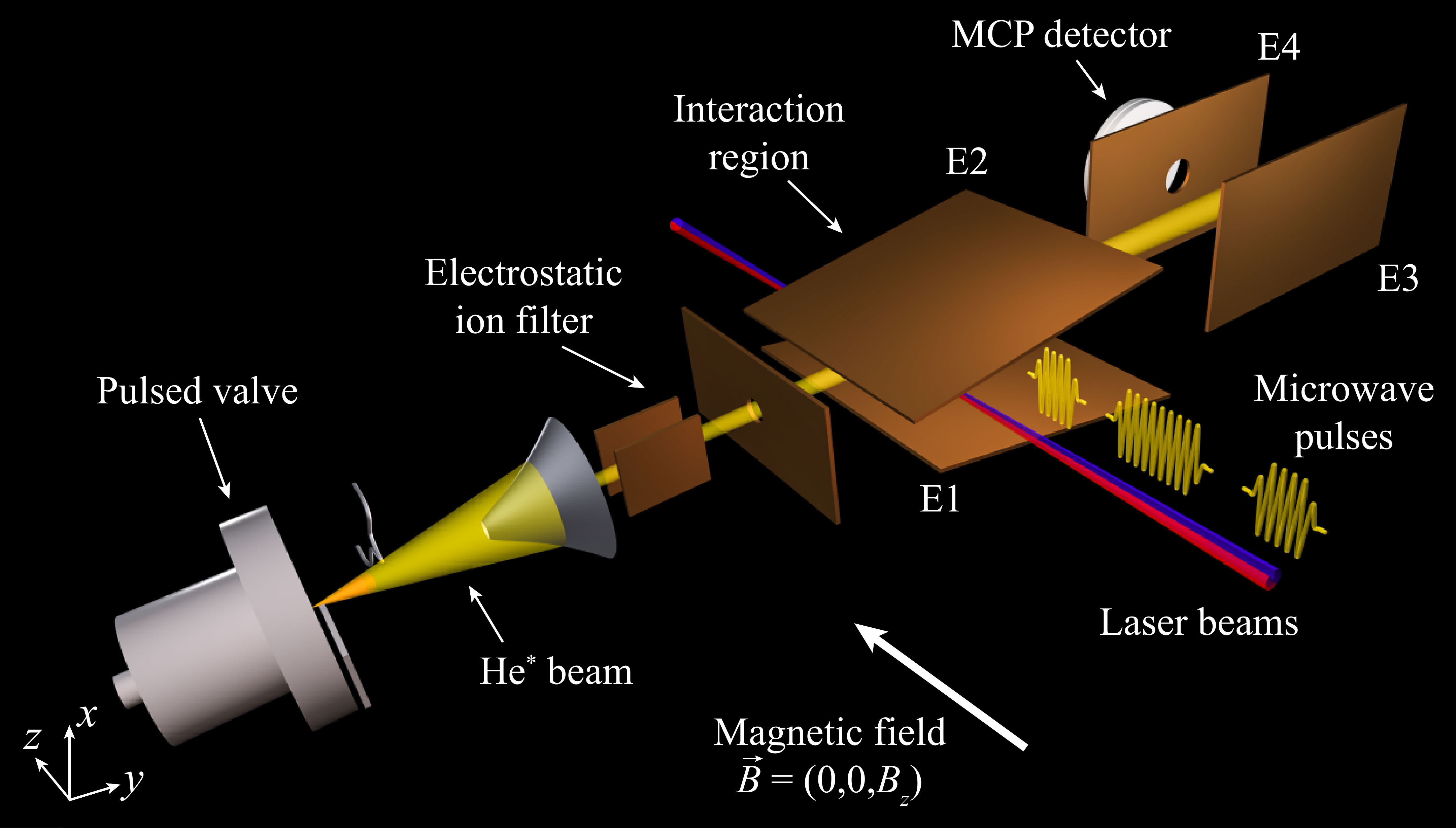}
\caption{Schematic diagram of the experimental apparatus (not to scale).}
\label{fig1}
\end{center}
\end{figure}

\begin{figure}
\begin{center}
\includegraphics[width = 0.4\textwidth, angle = 0, clip=]{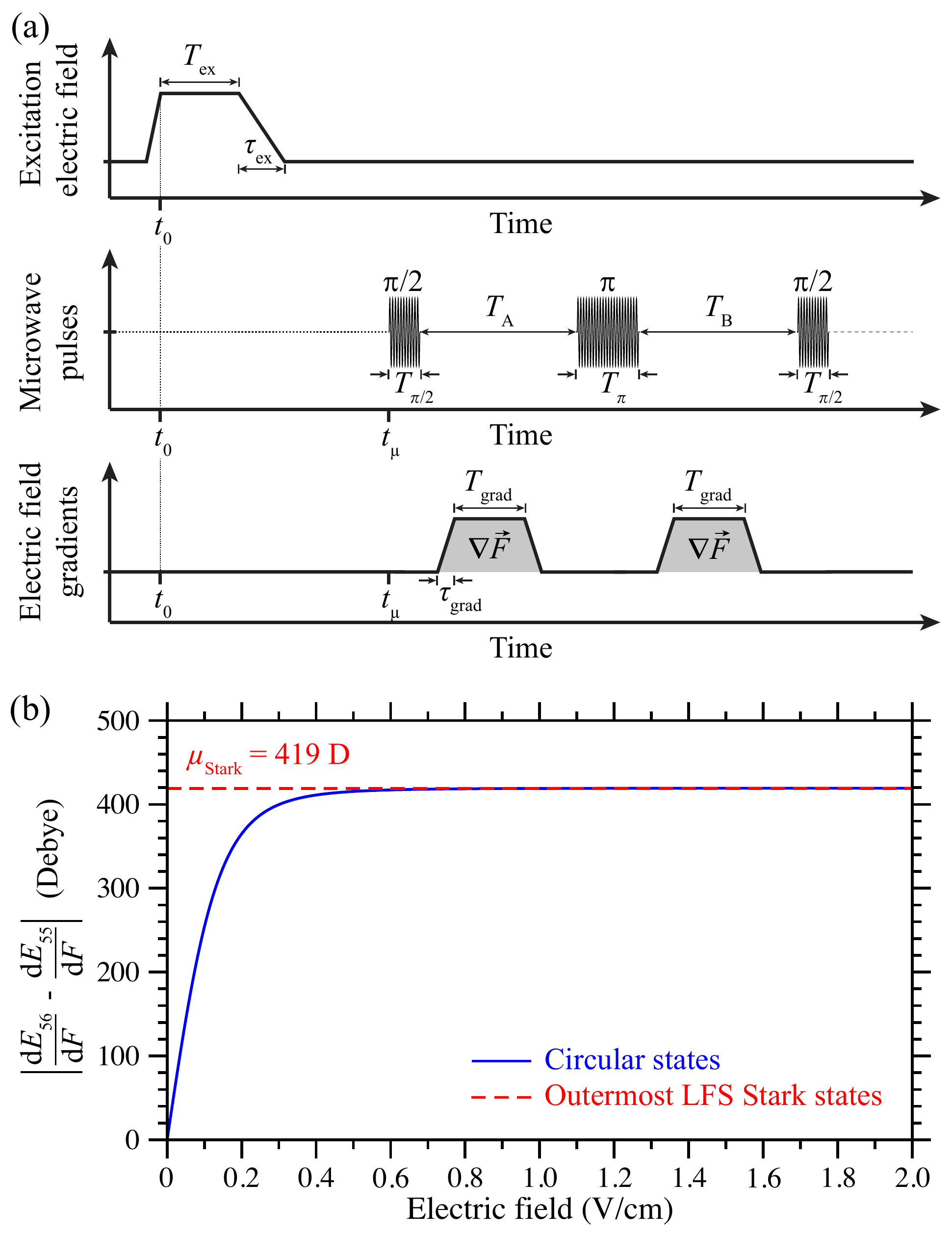}
\caption{(a) Sequence of electric field and microwave pulses. (b) Difference between the induced electric dipole moments, $\mathrm{d}E_{n\mathrm{c}}/\mathrm{d}F$, of the $|55\mathrm{c}\,\rangle$ and $|56\mathrm{c}\,\rangle$ states in He in an electric field applied perpendicular to a magnetic field $\vec{B} = (0,0,15.915)$~G. The difference between the static electric dipole moments of the outermost LFS Stark states in a pure electric field is indicated by the dashed red line.}
\label{fig2}
\end{center}
\end{figure}

A schematic diagram of the experimental apparatus is displayed in Fig.~\ref{fig1}. Metastable He atoms initially traveling in a pulsed (repetition rate 50~Hz) supersonic beam [mean velocity $\langle\vec{v}_0\rangle = (0,2000,0)$~m/s], were prepared in the $|n,\ell,m_{\ell}\rangle = |55,54,+54\rangle \equiv |55\mathrm{c}\rangle$ circular Rydberg state using the crossed-fields method~\cite{delande88a,zhelyazkova16a} ($\ell$ and $m_{\ell}$ are the orbital angular momentum and azimuthal quantum numbers, respectively). Circular state preparation was implemented within the wedge structure formed between the two $70\times105$~mm electrodes separated by 11.5~mm (35.8~mm) in the $x$ dimension at the end closest to the beam source (detection region), and labelled E1 and E2 in Fig.~\ref{fig1}. In this part of the apparatus the on-axis electric field gradient $\nabla\vec{F}_{\mathrm{grad}} = V_{\mathrm{grad}}\,(0,0.056,0)$~cm$^{-2}$ for a potential difference $V_{\mathrm{grad}}$ between E1 and E2. Rydberg state photoexcitation occurred at the position in the $y$ dimension where the separation between E1 and E2 was 19~mm, using a 1s2s\,$^3$S$_1\rightarrow$1s3p\,$^3$P$_2\rightarrow$1s$n$s/1s$n$d two-photon excitation scheme with focused (full-width-at-half-maximum beam waists $\sim100~\mu$m) co-propagating CW laser radiation~\cite{hogan18a}. To begin each experimental cycle a pulsed electric field, $\langle\vec{F}_{\mathrm{ex}}\rangle=(3.15,0,0)$~V/cm, was generated by applying a pulsed potential to E1 at time $t_0$ to tune the atomic energy levels into resonance with the lasers [see Fig.~\ref{fig2}(a)]. The electric field inhomogeneity of $\pm2$~mV/cm across the laser beams did not adversely affect circular state preparation. Photoexcitation was performed for $T_{\mathrm{ex}}=1.360~\mu$s in a static magnetic field $\vec{B} = (0,0,15.915)$~G. The excitation electric field was switched off in $\tau_{\mathrm{ex}}=1.0~\mu$s so that the excited atoms evolved adiabatically into the $|55\mathrm{c}\rangle$ state in zero electric field, with the $(32,0,0)$~mV/cm motional Stark effect cancelled~\cite{elliott95a}. Coherent population transfer between the $|55\mathrm{c}\rangle$ and $|56\mathrm{c}\rangle$ ($|56\mathrm{c}\rangle\equiv|56,55,+55\rangle$) states for interferometry was achieved using pulses of microwave radiation resonant with the $|55\mathrm{c}\rangle\rightarrow|56\mathrm{c}\rangle$ transition at 38.511\,313~GHz in the pure magnetic field (see Ref.~\cite{palmer19a} for further details). The implementation of $\pi/2$ ($\pi$) rotations on the Bloch sphere required pulse durations of $T_{\pi/2} \simeq 50$~ns ($T_{\pi}\simeq2T_{\pi/2}$) for the microwave intensities employed. At the end of each experimental cycle the excited atoms traveled for $60~\mu$s to the detection region between electrodes E3 and E4 where a slowly-rising pulsed electric field was generated for state-selective detection. The ionized electrons were collected on a microchannel plate (MCP) detector. 

Pairs of circular Rydberg states were used in these experiments because the large electric dipole moments for transitions between them ($\langle 56\mathrm{c}|e\hat{z}|55\mathrm{c} \rangle=5460$~D), and their low sensitivity to stray electric fields and electric field noise in electric fields near zero, allowed fast coherent state preparation and manipulation. In the experiments the decoherence times of these states, measured in zero electric field by Ramsey interferometry, were limited to $\sim5~\mu$s by the motion of the atoms. 

In perpendicular electric $\vec{F}=(F_x,0,0)$, and weak magnetic $\vec{B}=(0,0,B_z)$ fields, the energy shift, $E_{n\mathrm{c}}$, of a circular Rydberg state can be expressed as~\cite{pauli26a,delande88a}
\begin{eqnarray}
E_{n\mathrm{c}} &=& m_{\ell}\sqrt{\bigg(\mu_{\mathrm{B}} B_z\bigg)^2 + \bigg(\frac{3}{2}n\,e\,a\,F_x\bigg)^2},\label{eq:EStarkZeeman}
\end{eqnarray}
where $\mu_{\mathrm{B}}$ is the Bohr magneton, $e$ is the electron charge, and $a$ is the Bohr radius corrected for the reduced mass. In a constant magnetic field, circular states with $m_{\ell}>0$ have positive Stark shifts. In inhomogeneous electric fields, $\vec{F}_{\mathrm{grad}}$, atoms in these states experience a force, $\vec{f}=-\nabla E_{n\mathrm{c}} = -\nabla\vec{F}\,\mathrm{d}E_{n\mathrm{c}}/\mathrm{d}F$ ($\mathrm{d}E_{n\mathrm{c}}/\mathrm{d}F$ represents the induced electric dipole moment), toward regions of low field. They are therefore low field seeking (LFS). In weak (strong) electric fields, $E_{n\mathrm{c}}$ in Eq.~\ref{eq:EStarkZeeman} has an approximately quadratic (linear) dependence on $F_x$. In the case of the $|55\mathrm{c}\rangle$ and $|56\mathrm{c}\rangle$ states in He the induced dipole moments converge to 11\,320~D and 11\,740~D in strong fields. The interferometry scheme presented here relies on the state-dependent forces exerted on the two components of a coherent superposition of the $|55\mathrm{c}\rangle$ and $|56\mathrm{c}\rangle$ states that arise because of the difference in their electric dipole moments when $B_z=15.915$~G [see Fig.~\ref{fig2}(b)]. For $\langle\vec{F}_{\mathrm{grad}}\rangle<0.6$~V/cm this difference in dipole moments reflects the quadratic behaviour of the Stark shifts of the states. The maximal difference of 420~D occurs for $\langle\vec{F}_{\mathrm{grad}}\rangle>0.6$~V/cm and is equal to the difference in the dipole moments of the corresponding outer LFS Stark states in the absence of the magnetic field [dashed red curve in Fig.~\ref{fig2}(b)].


Rydberg-atom interferometry was implemented in 5 steps: 

\emph{Step 1}: At time $t_{\mu}=2.575~\mu$s [see Fig.~\ref{fig2}(a)], the atoms, with momentum $\vec{p}_0=4\,m_{\mathrm{u}}\,\langle\vec{v}_0\rangle$ ($m_{\mathrm{u}}$ is the atomic mass unit) were prepared in a coherent superposition of the $|55\mathrm{c},\vec{p}_0\,\rangle$ and $|56\mathrm{c},\vec{p}_0\rangle$ states by a $\pi/2$ pulse of resonant microwave radiation.

\emph{Step 2}: A pulsed inhomogeneous electric field, $\vec{F}_{\mathrm{grad}}$ (rise/fall time $\tau_{\mathrm{grad}}$; duration $T_{\mathrm{grad}}$) was then generated by applying a pulsed potential $V_{\mathrm{grad}}$ to electrode E2. The resulting state-dependent acceleration of the atoms, acting in the positive $y$ dimension, yielded a coherent superposition of the $|55\mathrm{c},\vec{p}_0+\mathrm{d}\vec{p}_{55\mathrm{c}}\,\rangle$ and $|56\mathrm{c},\vec{p}_0+\mathrm{d}\vec{p}_{56\mathrm{c}}\,\rangle$ states, where $\mathrm{d}\vec{p}_{n\mathrm{c}}$ is the state-dependent momentum change. The different momenta of these components resulted in a spatial separation that increased with time.

\emph{Step 3}: A $\pi$ pulse was then applied, in zero electric field, to invert the internal states of the two momentum components, i.e., $|55\mathrm{c},\vec{p}_0+\mathrm{d}\vec{p}_{55\mathrm{c}}\,\rangle\rightarrow|56\mathrm{c},\vec{p}_0+\mathrm{d}\vec{p}_{55\mathrm{c}}\,\rangle$ and $|56\mathrm{c},\vec{p}_0+\mathrm{d}\vec{p}_{56\mathrm{c}}\,\rangle\rightarrow|55\mathrm{c},\vec{p}_0+\mathrm{d}\vec{p}_{56\mathrm{c}}\,\rangle$. 

\emph{Step 4}: A second pulsed electric field gradient, of equal magnitude and duration to that in Step 2., was then applied. This resulted in the generation of a superposition of the $|55\mathrm{c},\vec{p}_0+\mathrm{d}\vec{p}_{56\mathrm{c}}+\mathrm{d}\vec{p}_{55\mathrm{c}}\,\rangle$ and $|56\mathrm{c},\vec{p}_0+\mathrm{d}\vec{p}_{55\mathrm{c}}+\mathrm{d}\vec{p}_{56\mathrm{c}}\,\rangle$ states. These states have equal momenta and hence de~Broglie wavelength, $\lambda_{\mathrm{dB}}$. However, they are spatially separated. 

\emph{Step 5}: A final $\pi/2$ pulse was applied at time $t_{\mu}+2.05~\mu$s, such that $T_{\mathrm{A}}=T_{\mathrm{B}}\simeq950$~ns [see Fig.~\ref{fig2}(a)], to interrogate the Rydberg state populations. 

The approach to state preparation and read-out used here is analogous to that  discussed in Ref.~\cite{borde89a} in the context of light-pulse atom interferometry. For $|V_{\mathrm{grad}}|=0$~V the contrast of the Ramsey fringes at the time of the final $\pi/2$ microwave pulse was $\sim0.8$.

\begin{figure}
\begin{center}
\includegraphics[width = 0.45\textwidth, angle = 0, clip=]{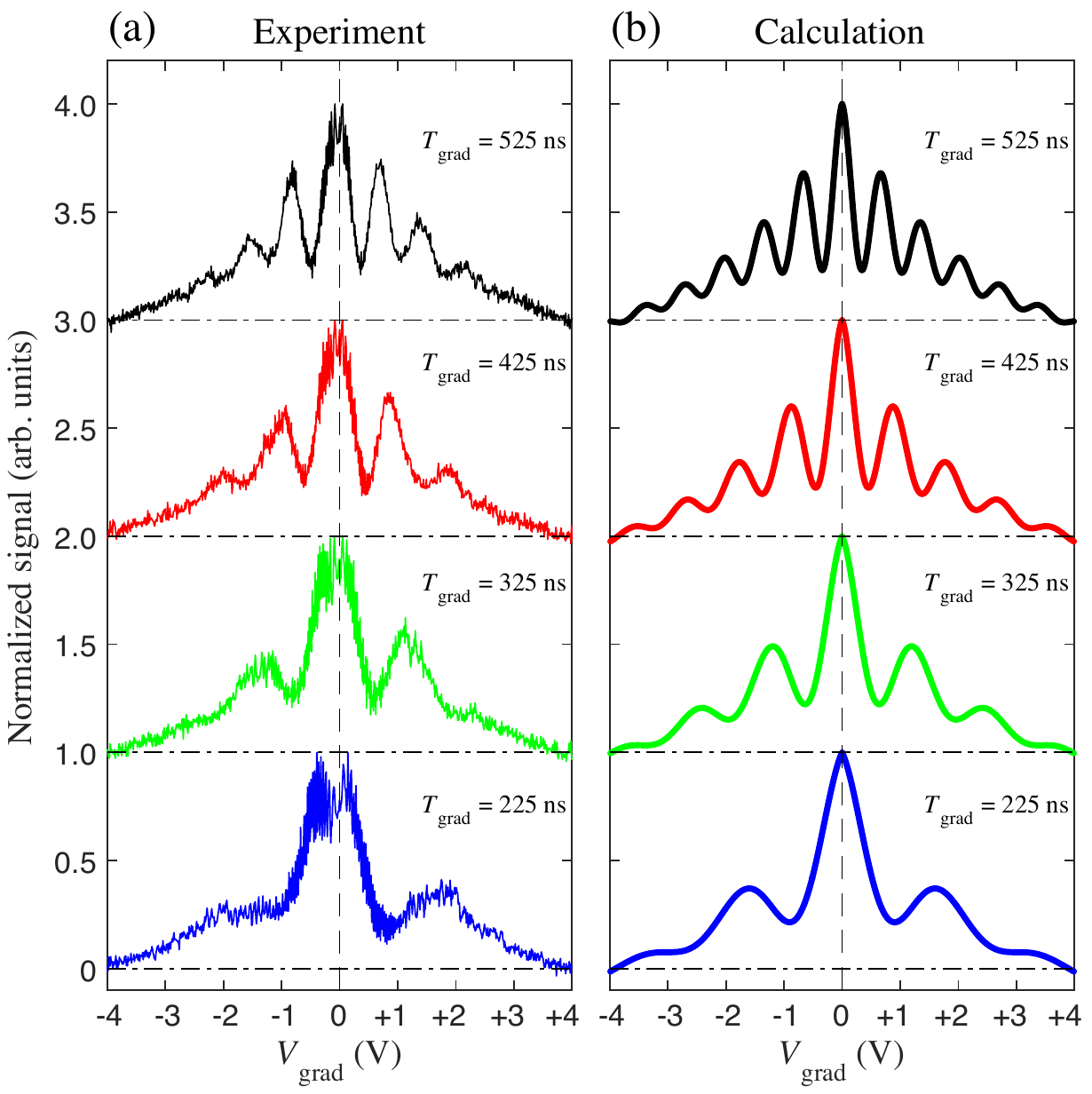}
\caption{(a) Measured, and (b) calculated Rydberg-atom interference patterns for $T_{\mathrm{grad}}=225$, 325, 425 and 525~ns (as indicated). Individual datasets are vertically offset for clarity.}
\label{fig3}
\end{center}
\end{figure}


In the experiments the ratio of the $|55\mathrm{c}\rangle$-to-$|56\mathrm{c}\rangle$ electron signals was monitored as $V_{\mathrm{grad}}$ was adjusted. In each dataset recorded the maximum value of this quantity was normalized to one. The results of measurements, for which $T_{\mathrm{grad}}=225$, 325, 425 and 525~ns, and $\tau_{\mathrm{grad}}=130$~ns, are displayed in Fig.~\ref{fig3}(a). The intensity maxima (minima) in the observed interference patterns correspond to conditions under which the matter waves were displaced in the $y$ dimension by integer (half integer) multiplies of $\lambda_{\mathrm{dB}}=h/p_0\simeq50$~pm ($p_0 = |\vec{p}_0|$) at the interrogation time. For $T_{\mathrm{grad}}=225$~ns, values of $|V_{\mathrm{grad}}|>1$~V were required to separate the matter waves by more than $\lambda_{\mathrm{dB}}/2$. Consequently, other than the intensity maximum at $V_{\mathrm{grad}}=0$~V, only one additional interference maximum was observed over the range of potentials applied. The sharp structure observed in this set of data for values of $|V_{\mathrm{grad}}|\lesssim0.5$~V are Ramsey fringes that arise from imperfections in the durations and amplitudes of the microwave and electric field pulses. For larger values of $T_{\mathrm{grad}}$, larger matter-wave displacements were achieved and additional interference fringes are seen. For $T_{\mathrm{grad}}=525$~ns the maximal displacement achieved was $\sim3\lambda_{\mathrm{dB}}=150$~pm. 

\begin{figure}
\begin{center}
\includegraphics[width = 0.45\textwidth, angle = 0, clip=]{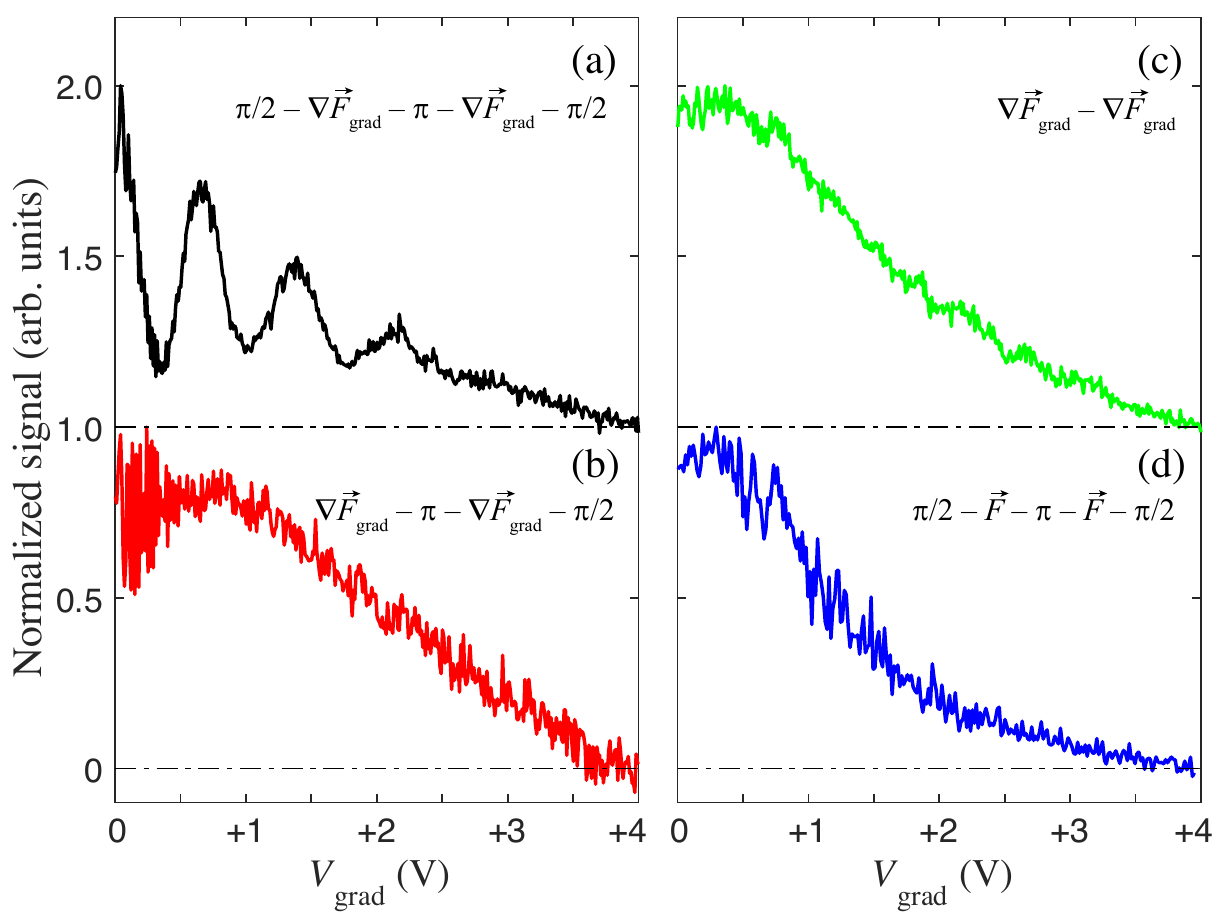}
\caption{Effects of modifications to the microwave and electric field pulse sequence on the interference patterns observed when $T=525$~ns. (a) The complete $\pi/2$\,--\,$\nabla\vec{F}_{\mathrm{grad}}$\,--\,$\pi$\,--\,$\nabla\vec{F}_{\mathrm{grad}}$\,--\,$\pi/2$ sequence as in Fig.~\ref{fig3}(a); (b) and (c) incomplete pulse sequences in which the first $\pi/2$ pulse was omitted, and only the pulsed electric field gradients were applied, respectively. (d) Parallel electrodes (see text for details). The data in (a) and (c) are vertically offset for clarity.}
\label{fig4}
\end{center}
\end{figure}

The results of semiclassical numerical calculations, performed to aid in the interpretation of the experimental data in Fig.~\ref{fig3}(a), are presented in Fig.~\ref{fig3}(b). These calculations involved integrating the classical equations of motion of atoms in the $|55\mathrm{c}\,\rangle$ and $|56\mathrm{c}\,\rangle$ states under the conditions for which the experiments were performed. The properties of the states were inverted at the time when the $\pi$ pulse was applied. At the end of each pulse sequence, the spatial separation between the trajectories was determined. The interference in the amplitudes of two plane waves with wavelength $\lambda_{\mathrm{dB}} = 50$~pm, and this spatial separation was then calculated. The resulting interference patterns in Fig.~\ref{fig3}(b) have the same periodicity as those in the experimental data. To achieve good quantitative agreement with the experimental data contributions from (1) decoherence, which reduced the contrast of the interference fringes, and (2) a loss in the total detected signal from circular states were included in the calculations. Both of these effects had an exponential dependence on the value of $|V_{\mathrm{grad}}|$, with decay constants of 1.2~V and 4.2~V, respectively (see Ref.~\cite{palmer19a} for further details). The decoherence results from a combination of the finite longitudinal coherence length associated with the momentum distribution of the Rydberg atoms, effects of electric field noise, and imperfections in the relative amplitudes of the electric field gradient pulses. 

To elucidate the mechanism by which population was lost as $|V_{\mathrm{grad}}|$ was increased, and to further validate the interpretation of the experimental data, additional measurements were performed, the results of which are presented in Fig.~\ref{fig4}. For clarity, only the parts of these datasets for which $V_{\mathrm{grad}}\geq0$~V are shown. The interference pattern in Fig.~\ref{fig4}(a) is a reference measurement and corresponds to that for $T_{\mathrm{grad}}=525$~ns in Fig.~\ref{fig3}(a). To confirm that an initial coherent superposition of Rydberg states was required to generate the observed interference patterns the first $\pi/2$ pulse was removed from the sequence in Fig.~\ref{fig2}(a). In this situation, see Fig.~\ref{fig4}(b), interference fringes are not visible. 

Next, a measurement was made with no microwave pulses applied. Interference fringes are also not visible in the resulting set of data in Fig.~\ref{fig4}(c). However, these data betray the origin of loss of population as $|V_{\mathrm{grad}}|$ was increased. The reduction in the signal in Fig.~\ref{fig4}(c) for larger values of $|V_{\mathrm{grad}}|$, combined with the requirements for adiabatic evolution of the $|55\mathrm{c}\rangle$ and $|56\mathrm{c}\rangle$ states in zero electric field, into the corresponding outer LFS Stark states in the combined magnetic and electric fields in the experiments~\cite{zhelyazkova16a}, leads to the conclusion that, for the constant value of $\tau_{\mathrm{grad}}=130$~ns used, the increase in the rate of change of the field experienced by the atoms for values of $|V_{\mathrm{grad}}|\gtrsim0.5$~V resulted in increasingly nonadiabatic dynamics when the pulsed fields were switched on and off. These nonadiabatic dynamics caused the loss of atoms from the circular states.

Finally, to confirm that the interference patterns in Fig.~\ref{fig3}(a) and Fig.~\ref{fig4}(a) resulted from the state-dependent forces on the atoms in the pulsed inhomogeneous electric fields, the experiments were repeated using an electrode configuration in which E1 and E2 were adjusted to lie parallel to each other with a separation of 13~mm. In this setup, the pulsed electric potentials did not give rise to significant electric field gradients at the position of the atoms. To allow comparison of the resulting data, in Fig.~\ref{fig4}(d), with the other datasets the electric potentials are scaled by 1.46 to account for the different separation between E1 and E2 in this configuration compared to that in the wedge configuration. The absence of interference fringes in this data further confirms the interpretation of the results in Fig.~\ref{fig3} and Fig.~\ref{fig4}(a).


In conclusion, we have demonstrated electric Rydberg-atom interferometry for the first time. The interferometer realized exploits the forces exerted on atoms with large electric dipole moments in inhomogeneous electric fields. The interference patterns observed resulted from differences in the forces acting on the two components of an initially prepared coherent superposition of Rydberg states of $\sim10^{-24}$~N. These forces caused relative accelerations of the matter wave components of $\sim100$~m/s$^{2}$. The maximal matter wave displacements measured were limited to $\sim3\lambda_{\mathrm{dB}}=150$~pm by the criteria for adiabatic evolution of the Rydberg states in the pulsed electric fields. In future experiments with circular Rydberg states, these requirements could be relaxed by modifying the combinations of magnetic and electric fields from those used here. Alternatively, experiments could be performed with lower $m_{\ell}$, or lower $\ell$ states for which adiabatic evolution in time-dependent electric fields can be more easily achieved. If through these refinements greater displacements between the matter wave components are achieved over longer timescales, the large spatial dimensions of the Rydberg atoms ($2\langle r\rangle=320$~nm) are expected to provide a new perspective from which to study spatial decoherence and wavefunction collapse in macroscopic quantum systems~\cite{bassi13a}. With modifications to the experimental configuration such that the forces induced by the electric field gradients are exerted on the atoms in the direction perpendicular to the surface of Earth, the methods presented here could be employed to measure the acceleration of Rydberg atoms in the gravitational field of Earth. To achieve a precision of 10\% in such measurements with the electrode configuration used here would require measurement times of $<100~\mu$s and matter-wave displacements of $<100$~nm. These timescales are well within the fluorescence lifetimes of the circular Rydberg states used here  ($\sim40$~ms), and low-$|m_{\ell}|$ Stark states in atomic hydrogen~\cite{seiler16a} or positronium~\cite{deller16a}. Measurements of this kind would open a new approach to test the weak equivalence principle with antihydrogen~\cite{kellerbauer08a,amole13a} or positronium~\cite{mills02a,cassidy14a} for which existing ground-state atom interferometric measurement techniques cannot be implemented.

\begin{acknowledgments}
We thank Dr. A. Deller (University College London) for valuable discussions. This work was supported by the Engineering and Physical Sciences Research Council under Grant No. EP/L019620/1, and the European Research Council (ERC) under the European Union's Horizon 2020 research and innovation program (Grant No. 683341).
\end{acknowledgments}
	

\end{document}